\documentclass[aps,prl,amsfonts,amssymb,twocolumn,amsmath,preprintnumbers,nofootinbib,floatfix,superscriptaddress]{revtex4-1}%
\usepackage{siunitx}
\usepackage{graphicx}
\usepackage{mathrsfs}
\usepackage{bm}
\usepackage{amsmath}
\usepackage{epstopdf}
\usepackage{dsfont}
\usepackage{amssymb}
\usepackage{tabularx}
\usepackage{array}
\usepackage{float}
\usepackage{color}
\usepackage{multirow}
\usepackage[T1]{fontenc}
\usepackage{makecell}
\usepackage[colorlinks,linkcolor=blue,anchorcolor=blue,citecolor=blue,urlcolor=blue]{hyperref}
\providecommand{\U}[1]{\protect\rule{.1in}{.1in}}

\newcommand{\vect}[1]{\boldsymbol{#1}}

\usepackage{amsfonts}
\usepackage{bm}
\usepackage{xspace}
\usepackage{dcolumn}
\usepackage{longtable}
\usepackage{multirow}
\usepackage{comment}
\def\llangle{\langle\hspace{-0.02cm}\langle}
\def\rrangle{\rangle\hspace{-0.02cm}\rangle}
\usepackage{pifont}
\newcommand{\cmark}{\ding{51}}
\newcommand{\xmark}{\ding{55}}

\makeatletter

\newcommand{\Rmnum}[1]{\expandafter\@slowromancap\romannumeral #1@}
\makeatother

\begin{document}

\title{Geometry-Driven Nonlinear Orbital Magnetoelectric Effect}
\author{Jinxiong Jia}
\affiliation{International Center for Quantum Design of Functional Materials and Department of Physics, University of Science and Technology of China, Hefei, Anhui 230026, China}
\affiliation{Hefei National Laboratory, University of Science and Technology of China, Hefei 230088, China}
\author{Zhenhua Qiao}
\email[]{qiao@ustc.edu.cn}
\affiliation{International Center for Quantum Design of Functional Materials and Department of Physics, University of Science and Technology of China, Hefei, Anhui 230026, China}
\affiliation{Hefei National Laboratory, University of Science and Technology of China, Hefei 230088, China}
\author{Jian Wang}
\email[]{jianwang@hku.hk}
\affiliation{College of Physics and Optoelectronic Engineering, Shenzhen University, Shenzhen 518060, China}
\affiliation{Quantum Science Center of Guangdong-Hongkong-Macao Greater Bay Area (Guangdong), Shenzhen 518045, China}
\affiliation{Department of Physics, The University of Hong Kong, Pokfulam Road, Hong Kong, China}

\begin{abstract}
    We propose a nonlinear orbital magnetoelectric effect, which generates orbital magnetization quadratically in centrosymmetric materials where the linear orbital magnetoelectric effect is strictly forbidden.
    Using extended semiclassical formulation, we derive a gauge-invariant microscopic theory that separates intrinsic and extrinsic contributions and establishes their distinct dependence on the relaxation time, providing an experimental discriminator.
    In two-dimensional systems the nonlinear response is far less constrained by out-of-plane rotational symmetries than the linear orbital magnetoelectric effect, substantially enlarging the materials platform. Microscopically, the dominant contributions are governed by a Hermitian-connection structure.
    Finally, we estimate that the magnitude of the nonlinear orbital magnetoelectric effect lies within the sensitivity of state-of-the-art magneto-optical Kerr measurements.
\end{abstract}

\maketitle

\noindent{\textit{\textcolor{blue}{Introduction}}}---
Orbitronics in condensed matter physics has attracted increasing attention in recent years~\cite{2024-Culcer-Adv_phys_X,2020-PRR-orbit_torque,2018-PRL-H.Lee,2025-NP-review_orbitronics,2005-PRL-SC_Zhang,2025-AEM-Yong_Jiang,2024-Johansson}.
Unlike spintronics, which manipulates the spin degree of freedom of Bloch electrons electrically~\cite{2019-RMP-SOT,2018-RMP-Antiferromagnet}, orbitronics explores generation and transport of orbital angular momentum, opening new pathways for information processing and storage~\cite{2025-npj_spintronics-Orbitronics}.
In particular, the orbital Hall effect, which generates a transverse flow of orbital angular momentum in response to an applied electric field, has been theoretically predicted and experimentally observed in various materials~\cite{2023-nature-OHE-Lee,2023-PRL-OHE-Gambardella,2023-PRL-OHE-Kawakami,2025-PRL-Extrinsic_OHE,2025-PRL-culcer,2024-PRL-Extrinsic_OHE,2024-PRL-Extrinsic_OHE-Culcer}.

However, similar to the spin Hall effect, the orbital Hall current is confined in-plane, resulting in orbital accumulation at the edges~\cite{2023-nature-OHE-Lee,2023-PRL-OHE-Gambardella,2023-PRL-OHE-Kawakami}.
An alternative approach is the orbital magnetoelectric effect (OME), which converts an applied electric field into a local orbital magnetization.
The extrinsic OME has been demonstrated to generate giant orbital torque, driving magnetization switching in twisted bilayer graphene~\cite{2019-science-twisted_graphene,2020-nc-KT_law,2021-PRL-MacDonald,2020-science-moire}, while the intrinsic OME has been recently formulated using both semiclassical theory and response theory~\cite{2021-PRB-Cong_Xiao,Culcer-2025-OME,2026-APL-haizhou-Lu}.
This provides a promising route to electrically control the orbital degree of freedom.
Nevertheless, since the orbital angular momentum is an axial vector, the OME requires inversion symmetry breaking and is therefore strictly forbidden in centrosymmetric materials.
In two-dimensional (2D) systems, the OME is further constrained by rotational symmetries.
Specifically, any rotational symmetry about the axis perpendicular to the 2D plane strictly forbids the OME~\cite{symmetry}, which significantly limits the number of candidate materials.

In this Letter, we propose a nonlinear orbital magnetoelectric effect (NOME), in which an orbital magnetization is induced quadratically by an applied electric field.
The NOME is characterized by a rank-3 pseudo-tensor $\chi_{c;ab}$, defined as $\delta M^c=\chi_{c;ab}E_aE_b$, where summation over repeated indices is implied.
As the quadratic field term $E_aE_b$ is even under inversion ($\mathcal{P}$), the NOME is permitted even in centrosymmetric materials.
Furthermore, in contrast to the linear OME, the NOME is far less constrained by rotational symmetries (see Table~\ref{tab1}), making it a promising platform for exploring orbital physics and converting orbital magnetization into spin-orbit torque in 2D materials.

Using an extended semiclassical formulation, we develop a microscopic theory that separates intrinsic and extrinsic contributions.
Owing to the singular nature of the position operator in periodic crystals~\cite{2000-PRB-Sipe,2025-PRL-culcer}, the NOME comprises three distinct components: a conventional contribution driven by orbital angular momentum matrix elements~\cite{2022-PRB-OHE_intra_inter_atom,2025-PRL-culcer}, which has a direct counterpart in nonlinear spin magnetization~\cite{Cong_xiao-PRL-2022}; and two geometric terms that are governed by the quantum geometric tensor and Hermitian connection.
We show that both the intrinsic and extrinsic NOME stem fundamentally from the Hermitian connection.

\bigskip
\noindent{\textit{\textcolor{blue}{Microscopic theory of NOME}}}---
Our analysis is based on the extended semiclassical theory~\cite{GaoY2014PRL,Cong_xiao-PRL-2022,2021-PRB-Cong_Xiao}, which has been successfully applied to a variety of nonlinear transport phenomena~\cite{2023-LJ_Xiang-Third_orderSemiclassical,2025-Jia-NMEE,2025-PRL-nonlinear_SHE} and is known to be equivalent to the quantum response theory~\cite{Jia2024}.
Within this framework, the orbital magnetization $\delta M^c$ is given by~\cite{SM}
\begin{align}
    \delta M^c=\int_{\rm BZ}\frac{{\rm d}\vect{k}}{(2\pi)^d}\left[g_n(\bar{\epsilon})\bar{\Omega}_n^c+\bar{f}_n(\bar{\epsilon})\langle \bar{W}|\hat{L}^c|\bar{W}\rangle\right].\label{eq1}
\end{align}
Here, $d$ is the spatial dimension and BZ the Brillouin zone.
$\hat{L}^c=\frac{1}{4}\varepsilon_{c\alpha\beta}\{\hat{r}^\alpha-r^\alpha_0,\hat{v}^\beta\}$ is the orbital angular momentum operator, where $r^\alpha_0$ denotes the center of mass of the wavepacket~\cite{2010-Niu-RMP}.
The state $|\bar{W}\rangle$ represents the normalized wavepacket expanded up to second order in the electric field~\cite{Jia2024}. The field induced Berry curvature is given by
$\bar{\Omega}_n^c=[\nabla\times (\mathcal{A}_n+a_n)]^c$, where $\mathcal{A}_n^\alpha=\langle u_{n\vect{k}}|i\partial_{k}^\alpha|u_{n\vect{k}}\rangle$ is the intraband Berry connection calculated from the periodic part of Bloch state $|u_{n\vect{k}}\rangle$ (satisfying $H_0(\vect{k})|u_{n\vect{k}}\rangle=\epsilon_{n\vect{k}}|u_{n\vect{k}}\rangle$), and $a_n^\alpha$ is the field-induced positional shift~\cite{Jia2024,GaoY2014PRL}. Furthermore,
$g_n=-k_BT\ln\left[1+e^{-(\bar{\epsilon}_n-\mu)/k_BT} \right]$ is the grand potential density, $\bar{f}_n$ is the non-equilibrium distribution function satisfying the Boltzmann equation, and $\bar{\epsilon}_n$ is the wave-packet energy up to second order in the electric field~\cite{Jia2024,2025-PRL-Definition_Nonlinear_Current,2022-PRB-Cong_Xiao-Third_Hall}.

The first and second terms in Eq.~\eqref{eq1} correspond to the revolution and self-rotation of the wave packet, respectively.
Combining the zeroth- and first-order wave-packets with the Boltzmann equation and Eq.~\eqref{eq1}, recovers the equilibrium orbital magnetization and linear OME, respectively; the latter includes both intrinsic and extrinsic contributions~\cite{SM}, in agreement with previous findings~\cite{Culcer-2025-OME,2021-PRB-Cong_Xiao,2016-PRL-JEMoore,2020-PRB-Cong_Xiao,2010-Niu-RMP,2005-PRL-DiXiao,2007-PRL-JShi,2005-Resta-PRL,2015-Yoda}.
Finally, by substituting the second-order wave-packet~\cite{Jia2024} and the expanded distribution function $f(\bar{\epsilon})$ into Eq.~\eqref{eq1}, we find that the intrinsic NOME comprises three distinct contributions:
\begin{align}
    \chi_{c;ab}^{(0)}=\chi_{c;ab}^{(0,od)}+\chi_{c;ab}^{(0,d)}+\chi_{c;ab}^{(0,ic)}.
\end{align}
Here, "od" denotes the conventional contribution from the  matrix elements of orbital angular momentum operator, $L_{nm}^c=\frac{1}{4}\varepsilon_{c\alpha\beta}\sum_\ell\{r^\alpha,v^\beta\}_{nm}$. In this expression, $r_{nm}^a=\langle u_{n\vect{k}}|i\partial_k^a|u_{m\vect{k}}\rangle$ ($n\neq m$) is the interband Berry connection matrix, and $v_{nm}^a=\delta_{nm}\partial_k^a \epsilon_{n}+i\epsilon_{nm}r_{nm}^a$ is the velocity operator matrix with $\epsilon_{nm}=\epsilon_n-\epsilon_m$. The terms "d" and "ic" refer to the dipole-type and positional shift geometric contributions, respectively.
Since these geometric terms originate solely from the diagonal part of position operator~\cite{2025-PRL-culcer,SM}, they are naturally described by the Hermitian connection, a gauge-covariant interband geometric structure that governs nonlinear responses beyond the quantum geometric tensor by encoding the local unitary frame rotation of Bloch eigenstates across momentum space. This connection is defined as $C_{nm}^{\alpha ab}=-iv^\alpha_{nm}\mathcal{D}_{mn}^a r^b_{mn}/\epsilon_{nm}=\mathcal{C}_{nm}^{\alpha ab}-i\mathcal{F}_{nm}^{\alpha ab}/2$~\cite{2022-NP-nagaosa}. 
Similarly, quantum geometric tensor is given by $g_{nm}^{ab}=-iv^a_{nm}r^b_{mn}/\epsilon_{nm}=\mathcal{G}_{nm}^{ab}-i\Omega_{nm}^{ab}/2$~\cite{2025-Xiang-PRL-Gyrotropic,2022-NP-nagaosa,2025-Nagaosa-review}. In these expressions, $\mathcal{D}_{mn}^a=\partial_k^a-i(\mathcal{A}_m^a-\mathcal{A}_n^a)$ is the covariant derivative, $\mathcal{C}_{nm}^{\alpha ab}$ is the metric connection, $\mathcal{F}_{nm}^{\alpha ab}$ is the symplectic connection, and $\mathcal{G}_{nm}^{ab}$ and $\Omega_{nm}^{ab}$ are the local quantum metric and Berry curvature, respectively.
Using these definitions, the dominant intrinsic NOME takes the form~\cite{SM}
\begin{align}
    \chi_{c;ab}^{(0,od)}&=\sum_{nm}\left[f_n\left(\bar{L}_{mn}^c\frac{\mathcal{G}^{ab}_{nm}}{\epsilon_{nm}^2}+2\bar{v}^a_{nm}\frac{\bar{\mathcal{G}}_{nm}^{cb}}{\epsilon_{nm}^2}-2\frac{\bar{\mathcal{C}}_{nm}^{cab}}{\epsilon_{nm}} \right)\right.\notag\\
    &\hspace{1.2cm}\left.-f_n'\frac{\mathcal{G}_{nm}^{ab}L_n^c}{\epsilon_{nm}}\right], \label{chi_od}\\
    \chi_{c;ab}^{(0,ic+d)}&=\frac{\varepsilon_{c\alpha\beta}}{4}\sum_{nm}f_n\left[(v_n^\beta+v_m^\beta) \mathcal{N}_{nm}^{\alpha ab} +\nabla_\beta \mathcal{L}_{nm}^{\alpha ab} \right],\label{chi_IC}
\end{align}
where $f_n$ is the Fermi-Dirac distribution function, $f_n'=\partial f_n/\partial \epsilon_n$, $\bar{v}_{nm}^a=\partial_k^a\epsilon_{nm}$, and $\bar{L}_{mn}^c=L_{mm}^c-L_{nn}^c$. 
The components of Eq.~\eqref{chi_IC} are defined as
\begin{align}
    \mathcal{N}_{nm}^{\alpha ab}&=\frac{\mathcal{F}_{nm}^{\alpha ab}+\mathcal{F}_{nm}^{a\alpha b}}{\epsilon_{nm}^2}+\bar{v}_{mn}^b\frac{\Omega_{nm}^{\alpha a}}{\epsilon_{nm}^3},\\
    \mathcal{L}_{nm}^{\alpha ab}&=\frac{\mathcal{F}_{nm}^{\alpha ab}}{\epsilon_{nm}}+\frac{\Omega_{nm}^{\alpha a}\bar{v}^b_{mn}}{\epsilon_{nm}^2}.
\end{align}
We further define $\bar{g}_{nm}^{ab}=-iL_{nm}^ar_{mn}^b/\epsilon_{nm}=\bar{\mathcal{G}}_{nm}^{ab}-i\bar{\Omega}_{nm}^{ab}/2$ and $\bar{\mathcal{C}}_{nm}^{cab}=-{\rm Re}\left[iL_{nm}^c\mathcal{D}_{mn}^ar_{mn}^b\right]/\epsilon_{nm}$.
These quantities share the same structural form as $g_{nm}^{ab}$ and $\mathcal{C}_{nm}^{\alpha ab}$, except that the velocity operator is replaced by the orbital angular momentum operator.

Furthermore, substituting the nonequilibrium distribution function $f_n^{(1)}=\tau E_a\partial_a f_n$ into Eq.~\eqref{eq1} yields the extrinsic NOME contribution, $\chi_{c;ab}^{(1)}=\chi_{c;ab}^{(1,od)}+\chi_{c;ab}^{(1,ic)}+\chi_{c;ab}^{(1,d)}$, with its components defined as
\begin{alignat}{2}
    &\chi_{c;ab}^{(1,od)}&&=-\tau\sum_{nm} f_n\partial^a \bar{\Omega}_{nm}^{cb},\label{chi_1_od}\\
    &\chi_{c;ab}^{(1,ic)}&&=-\tau\varepsilon_{c\alpha\beta}\sum_n f_n \partial^a \left[\frac{\mathcal{G}_{nm}^{\alpha b}v_n^\beta}{\epsilon_{nm}}\right],\label{chi_1_IC}\\
    &\chi_{c;ab}^{(1,d)}&&=\frac{\tau}{2}\varepsilon_{c\alpha\beta}\sum_{nm}f_n\partial^a\left[\frac{\bar{v}_{nm}^\beta \mathcal{G}^{\alpha b}_{nm}}{\epsilon_{nm}}+\mathcal{C}_{nm}^{\beta\alpha b}\right].\label{chi_1_d}
\end{alignat}
Similarly, "ic", "d", and "od" represent the itinerant circulation, dipole-type local circulation, and conventional local circulation terms, respectively.

Equations \eqref{chi_od}-\eqref{chi_IC} and \eqref{chi_1_od}-\eqref{chi_1_d} constitute the central results of this Letter.
Several important remarks are in order.
First, the distinct scaling behaviors provide a direct experimental discriminator: the intrinsic response $\chi_{c;ab}^{(0)}$ includes both Fermi-sea and Fermi-surface contributions, whereas $\chi_{c;ab}^{(1)}$ is a purely Fermi-surface term that scales linearly with $\tau$. This dependence allows one to separate the contributions via disorder engineering or mobility control.
Second, the geometric contributions $\chi_{c;ab}^{(j,ic)}$ and $\chi_{c;ab}^{(j,d)}$ for $j=0,1$ are unique to orbital magnetization, having no counterparts in the nonlinear spin magnetization.
Finally, by replacing $L_{nm}^c$ in $\chi_{c;ab}^{(j,od)}$ with the velocity $v_{nm}^c$, spin $\sigma_{nm}^c$ or spin-current $(J^\lambda_c)_{mn}=\frac{1}{2}\{\hat{s}^\lambda,\hat{v}^c\}_{mn}$ matrices, one reproduces the established formulas for nonlinear Hall effect~\cite{GaoY2014PRL,2015-PRL-LiangFu}, nonlinear spin magnetization~\cite{Cong_xiao-PRL-2022,2023-PRL-Xiao_C}, and nonlinear spin Hall effect~\cite{2025-PRL-nonlinear_SHE,2025-Xiang-PRL-spin} for order $\tau^j$, respectively.

\begin{table*}[t]
\centering
\renewcommand{\arraystretch}{1.5}
\caption{Constraints on the intrinsic NOME ($\chi_{z;ab}^{(0)}$) and extrinsic NOME ($\chi_{z;ab}^{(1)}$) imposed by various magnetic point group symmetries in 2D systems.
The symbols \cmark (\xmark) indicate whether the corresponding NOME component is symmetry-allowed (forbidden), respectively.
Here, the system is assumed to lie in the $xy$ plane, and the first four columns corresponds to symmetries that forbid the linear OME.
}
\label{tab1}
\begin{tabularx}{17cm}{>{\centering\arraybackslash}m{1cm} *{4}{>{\centering\arraybackslash}m{1.9cm}}|>{\centering\arraybackslash}m{1.9cm}>{\centering\arraybackslash}m{3cm}>{\centering\arraybackslash}m{2cm}}
    \hline\hline
    &$\mathcal{P},C_{2}^z,\mathcal{M}^z$&$C_{3,4,6}^z,S_{3,4,6}^z$&$C_{3,6}^z\mathcal{T},S_{3,6}^z\mathcal{T}$&$C_{4}^z\mathcal{T},S_{4}^z\mathcal{T}$&$C_{2}^{x,y},\mathcal{M}^{x,y}$&$\mathcal{T},\mathcal{P}\mathcal{T},C_2^z\mathcal{T},\mathcal{M}^z\mathcal{T}$&$C_2^{x,y}\mathcal{T},\mathcal{M}^{x,y}\mathcal{T}$\\
    \hline
    $\chi_{z;xx}^{(0)}$&\cmark&$\chi_{z;yy}^{(0)}$&\xmark&$-\chi_{z;yy}^{(0)}$&\xmark&\xmark&\cmark\\
    $\chi_{z;xy}^{(0)}$&\cmark&\xmark&\xmark&\cmark&\cmark&\xmark&\xmark\\
    $\chi_{z;yy}^{(0)}$&\cmark&\cmark&\xmark&\cmark&\xmark&\xmark&\cmark\\
    $\chi_{z;xx}^{(1)}$&\cmark&$\chi_{z;yy}^{(1)}$&$\chi_{z;yy}^{(1)}$&$\chi_{z;yy}^{(1)}$&\xmark&\cmark&\xmark\\
    $\chi_{z;xy}^{(1)}$&\cmark&\xmark&\xmark&\xmark&\cmark&\cmark&\cmark\\
    $\chi_{z;yy}^{(1)}$&\cmark&\cmark&\cmark&\cmark&\xmark&\cmark&\xmark\\
    \hline\hline
\end{tabularx}
\end{table*}

\bigskip
\noindent{\textit{\textcolor{blue}{Symmetry property}}}---
As a rank-3 pseudo-tensor, $\chi_{c;ab}^{(0/1)}$ is $\mathcal{P}$-even. Furthermore, $\chi_{c;ab}^{(0)}$ is $\mathcal{T}$-odd, while $\chi_{c;ab}^{(1)}$ is $\mathcal{T}$-even.
Unlike spin magnetization or the spin Hall effect, which rely directly on spin-orbit coupling (SOC), the intrinsic NOME involves only orbital operators and might initially appear independent of SOC.
However, in the absence of SOC, the system usually possesses an additional spin group symmetry $[\bar{C}_2||\mathcal{T}]$~\cite{2022-PRX-Jungwirth} that constrains the intrinsic NOME to zero~\cite{note1}.
By contrast, the extrinsic NOME persists even without SOC~\cite{2024-npj_spintronics,2026-PRB-Lu,2026-PRL-Lu} due to its $\mathcal{T}$-even nature.
This analysis reveals that the intrinsic and extrinsic NOME obey the same symmetry constraints as the intrinsic anomalous Hall effect and intrinsic orbital Hall effect.
More generally, constraints imposed by arbitrary magnetic point group (MPG) symmetries are dictated by the transformation rule~\cite{Newnham}
\begin{align}
    \chi_{c;ab}^{(0/1)}=(\eta_T)^{1/0}\, {\rm det}(R)R_{aa'}R_{bb'}R_{cc'}\chi_{c';a'b'}^{(0/1)},
\end{align}
where $R_{aa'}$ denotes the matrix representation of the MPG symmetry, and $\eta_T=-1$ ($1$) for MPG operations that include (do not include) time-reversal.
Here, we focus on 2D systems, where the magnetization arising from the angular Hall current vanishes.
Table~\ref{tab1} summarizes the constraints imposed by various MPG symmetries on the intrinsic and extrinsic NOME components; notably, the first four columns correspond to symmetries that forbid the linear OME.
As a result, we identify 41 and 55 MPGs that support the intrinsic and extrinsic NOME in 2D systems, respectively~\cite{SM}.
In stark contrast, the intrinsic and extrinsic OMEs are allowed only in 10 and 8 MPGs, respectively~\cite{SM}.
Thus, NOME provides a promising route to generating orbital magnetization in a significantly broader range of materials than the linear OME.

In the following, we investigate two systems: one exhibiting only intrinsic NOME and the other allowing solely extrinsic NOME. 
We show that in both cases, the geometric contribution dominates the conventional contribution.

\bigskip
\noindent\textcolor{blue}{Honeycomb lattice model}---
We illustrate the intrinsic NOME using the modified Kane-Mele model, defined as~\cite{2005-PRL-kane_mele,2014-PRL-qiao-Valley_Polar_QAH,Liu-feng,S_Hasegawa}
\begin{align}
    H=&-t\sum_{\langle ij\rangle}c_{i\alpha}^\dag c_{j\alpha}+i\lambda_R\sum_{\langle ij \rangle}c_{i\alpha}^\dag(\vect{\sigma}_{\alpha\beta}\times \vect{d}_{ij})_zc_{j\beta}\notag\\
    &+i\lambda_{so}\sum_{\llangle ij\rrangle} v_{ij}c_{i\alpha}^\dag \sigma^z_{\alpha\beta}c_{j\beta}+\lambda\sum_{i}c_{i\alpha}^\dag s^z_{\alpha\beta}c_{i\beta}.\label{Kane-Mele}
\end{align}
Here, $t$ denotes the nearest-neighbor hopping, the second term describes the inversion-breaking Rashba-type SOC, the third term corresponds to the mirror SOC, and the last term represents the $\mathcal{T}$-breaking exchange coupling.
The system possesses $6m'm'$ MPG symmetry (containing $C_{6z}$ and $\mathcal{T}M_x$).
Consequently, Table~\ref{tab1} indicates that only the intrinsic NOME component $\chi_{z;xx}^{(0)}=\chi_{z;yy}^{(0)}$ is finite, while both the extrinsic NOME and linear OME vanish.

\begin{figure}
    \centering
    \includegraphics[width=1\linewidth]{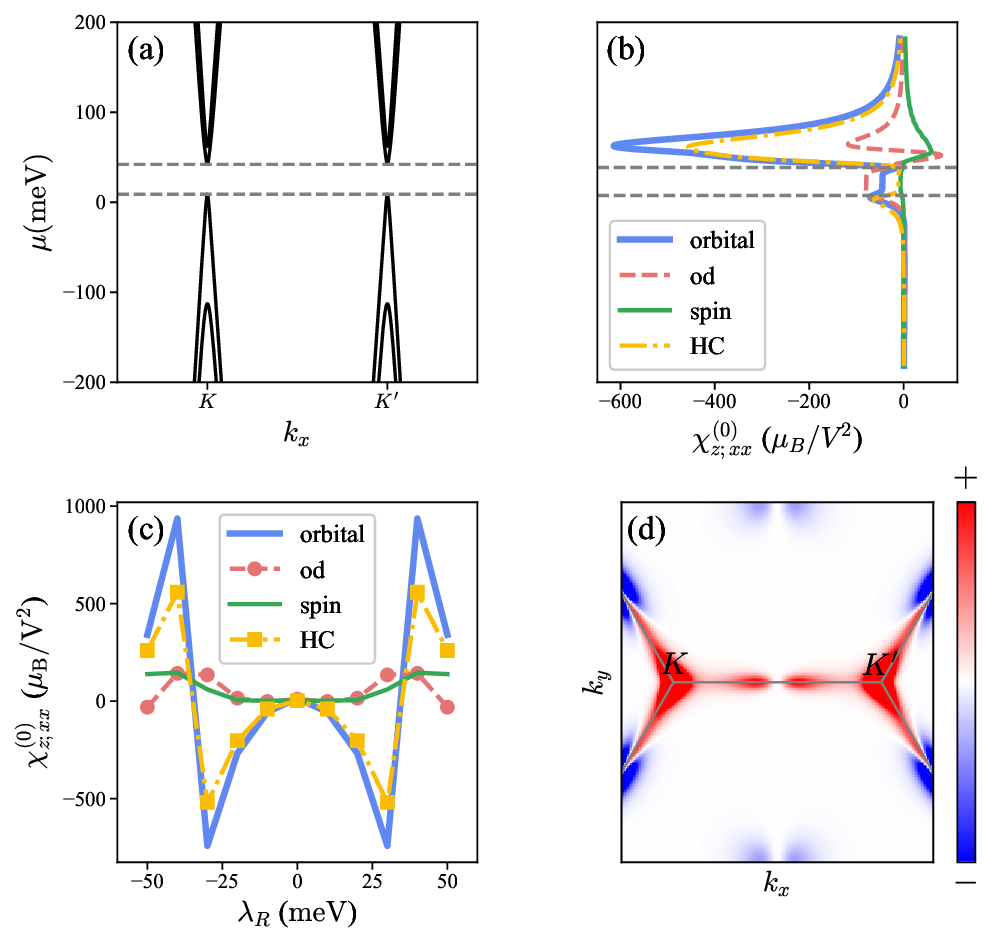}
    \caption{
    (a) Band dispersion for Eq.~\eqref{Kane-Mele}. (b) The intrinsic NOME component $\chi_{z;xx}^{(0)}$, its three contributions as a function of chemical potential $\mu$.
    For comparison, we also plot the nonlinear spin magnetization component $\chi_{z;xx}^{s(0)}$, and the Hermitian connection of $\chi_{z;xx}^{(0)}$ termed as HC.
    (c) The Rashba SOC $\lambda_R$ dependence of intrinsic NOME $\chi_{z;xx}^{(0)}$ at fixed $\mu=50{\rm meV}$. (d) The $\vect{k}$-resolved distribution of the integrand of $\chi_{z;xx}^{(0,d)}+\chi_{z;xx}^{(0,ic)}$ for the third band.
    Parameters: $t=0.85{\rm eV}$, $\lambda_{R}=20{\rm meV}, \lambda=10{\rm meV}$, $\lambda_{so}=10{\rm meV}$, and $T=20{\rm K}$.}
    \label{kane-mele_fig}
\end{figure}

\begin{figure}[t]
    \centering
    \includegraphics[width=1\linewidth]{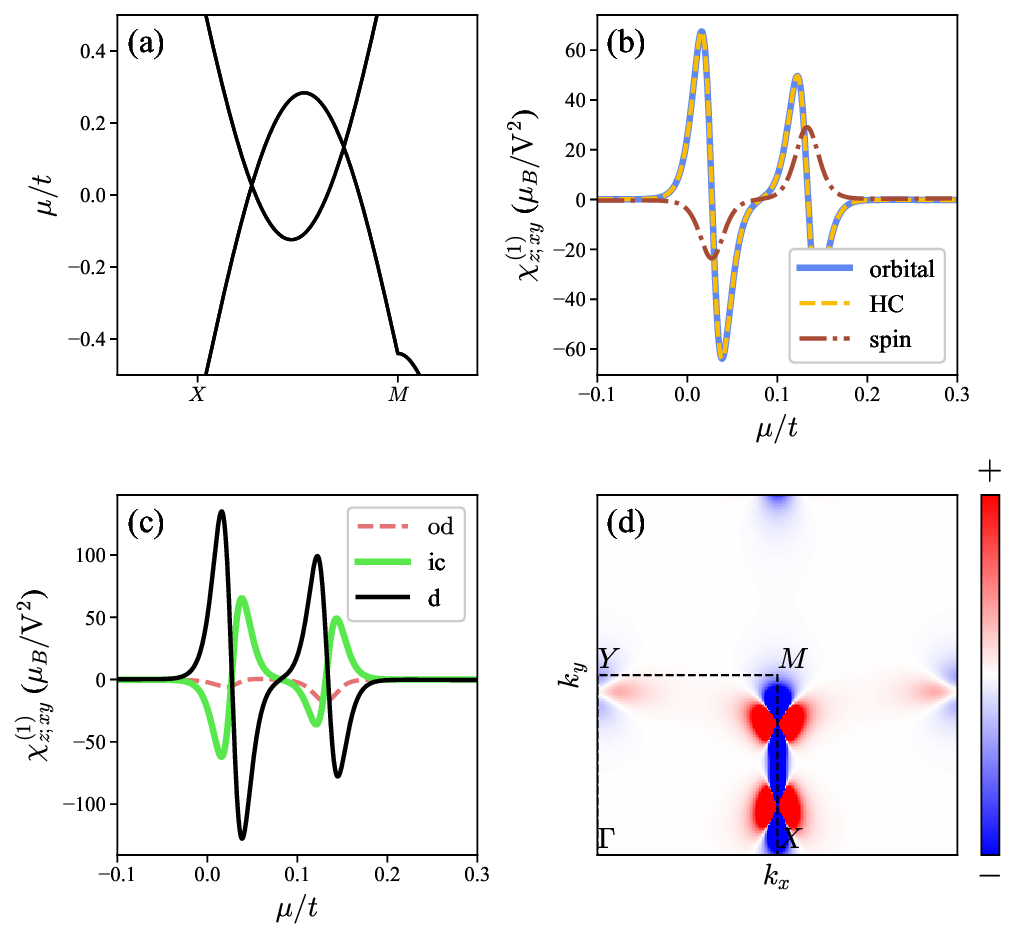}
    \caption{(a) Band dispersion for Eq.~\eqref{CuMnAs}.
    (b) The extrinsic nonlinear magnetization including orbital and spin components as a function of chemical potential $\mu$.
    (c) The three components of extrinsic nonlinear orbital magnetization $\chi_{z;xy}^{(1)}$ as a function of chemical potential $\mu$.
    (d) The $\vect{k}$-resolved distribution of the integrand of $\chi_{z;xy}^{(1,d)}$ for first two bands.
    Here, we use $t'=0.08t$, $J_n=0.6t$, $\lambda=0.8t$, $T=100{\rm K}$ and $\tau=10{\rm fs}$.
    }
    \label{Fig_CuMnAs}
\end{figure}

Fig.~\ref{kane-mele_fig}(a) displays the band dispersion of Eq.~\eqref{Kane-Mele} using the parameters given in the caption.
In Fig.~\ref{kane-mele_fig}(b), we plot the intrinsic NOME component $\chi_{z;xx}^{(0,od)}$ and total $\chi_{z;xx}^{(0)}$ as a function of chemical potential $\mu$, alongside the nonlinear spin magnetization for comparison.
We observe that the geometric "d" and "ic" contributions dominate the intrinsic NOME when $\mu$ lies near conduction band edge.
Notably, the Hermitian connection contribution further dominates over these geometric contributions.
When all orbital components are included, the generated orbital magnetization is approximately three times larger than the spin magnetization and carries the opposite sign.
Fig.~\ref{kane-mele_fig}(d) illustrates the $\vect{k}$-resolved integrand of $\chi_{z;xx}^{(0,d)}+\chi_{z;xx}^{(0,ic)}$ for the third band; this map reveals that the peak response originates primarily from the region around the band edges at $K$ and $K'$ points. Finally,
Fig.~\ref{kane-mele_fig}(c) presents dependence of intrinsic NOME $\chi_{z;xx}^{(0)}$ on the Rashba SOC strength at fixed $\mu=50{\rm meV}$.
At $\lambda_R=0$, where the inversion symmetry is restored, the intrinsic NOME retains a non-zero minimum of $\sim 10\mu_B/V^2$ and $\chi_{z;xx}^{(0)}$ behaves as an even function of $\lambda_R$, consistent with its $\mathcal{P}$-even symmetry.

\bigskip
\noindent\textcolor{blue}{\textit{Antiferromagnets}}---
To illustrate the extrinsic NOME, we consider the minimal $\mathcal{P}\mathcal{T}$-symmetric model of the antiferromagnet CuMnAs, which features two magnetic sublattices related by $\mathcal{P}\mathcal{T}$ symmetry.
Its tight-binding Hamiltonian reads~\cite{2017-PRL-Jungwirth}
\begin{align}
    H=&-2t\hat{\tau}_x\cos\frac{k_x}{2}\cos\frac{k_y}{2}-t'(\cos k_x+\cos k_y)\notag\\
    &+\lambda\hat{\tau}_z(\hat{\sigma}_y\sin k_x-\hat{\sigma}_x\sin k_y)+J_n\hat{\tau}_z\hat{\sigma}_x,\label{CuMnAs}
\end{align}
where $\hat{\tau}$ and $\hat{\sigma}$ are Pauli matrices for the sublattice and spin sectors, respectively. $t$ ($t'$) is the nearest (next-nearest) neighbor hopping, $\lambda$ is the Rashba-type SOC strength, and $J_n$ is the exchange coupling.
This system preserves $\mathcal{P}\mathcal{T}$, $C_{2z}\mathcal{T}$ and $C_{2y}$ symmetries, with both $\mathcal{P}\mathcal{T}$ and $C_{2y}$ interchanging the magnetic sublattices.
Consequently, while $\mathcal{P}\mathcal{T}$ symmetry forbids the intrinsic NOME, the $C_{2y}$ symmetry allows a single non-zero extrinsic NOME component, $\chi_{z;xy}^{(1)}$, consistent with Table~\ref{tab1}.

Figure~\ref{Fig_CuMnAs}(a) illustrates the band dispersion, highlighting two Dirac points along the high-symmetry path X-M.
In Fig.~\ref{Fig_CuMnAs}(b), we plot the extrinsic nonlinear magnetization versus $\mu$; notably, the orbital component dominates over the spin component. 
We further find that the Hermitian connection contribution, remains the dominant term in the extrinsic NOME throughout the entire range of $\mu$.The $\vect{k}$-resolved map of the $\chi_{z;xy}^{(1,d)}$ integrand for the first two bands [Fig.~\ref{Fig_CuMnAs}(d)], confirms that the main contribution comes from the Dirac points, reflecting the interband nature of the NOME.
In addition, we plot the three contributions of extrinsic nonlinear orbital magnetization $\chi_{z;xy}^{(1)}$ versus $\mu$ in Fig.~\ref{Fig_CuMnAs}(c).

\bigskip
\noindent\textcolor{blue}{\textit{Discussion and conclusion}}---
We assess feasibility by estimating the NOME magnitude in our models.
Under a moderate electric field $E=10^5{\rm V/m}$ along the $x$-axis, the honeycomb lattice yields an induced orbital magnetization of $\sim 10^{-5}\mu_B/{\rm nm}^2$ [Fig.~\ref{kane-mele_fig}(c)], while
CuMnAs yields $\sim 10^{-6}\mu_B/{\rm nm}^2$ [Fig.~\ref{Fig_CuMnAs}(b)].
Both values are sufficiently large to be detected via polar magnetooptical Kerr effect experiments~\cite{2023-PRL-OHE-Kawakami,2023-PRL-OHE-Gambardella,2026-PRB-Lu}.

Although the net $\chi_{c;ab}^{(0)}$ vanishes in $\mathcal{P}\mathcal{T}$-symmetric materials, a staggered orbital magnetization can still emerge on different sublattices, analogous to the extrinsic OME in antiferromagnets~\cite{2019-nc-OEE-antiferromagnet}.
Notably, this staggered intrinsic NOME generally requires lower symmetry than the extrinsic OME.
For instance, in a system with $\mathcal{T}\tau_{1/2}$ symmetry, where $\tau_{1/2}$ represents the half-lattice translation, the staggered NOME is allowed, while the linear counterpart is forbidden.
It should be emphasized that the NOME does not directly generate spin-orbit torque.
Instead, the SOC is required to convert the orbital magnetization into a spin accumulation, thereby exerting a torque on the magnetic order parameter, similar to the orbital torque mechanism~\cite{2020-PRR-orbit_torque}.
While we focus on the $\tau^0$ and $\tau^1$ contributions within the relaxation time approximation, an additional extrinsic contribution, referred to as the Drude NOME, arises at order $\tau^2$
\begin{align}
    \chi_{c;ab}^{(2)}=&\tau^2\sum_{n}\left[\Omega_n^c\partial_k^{ab}g_n+L_n^c\partial_k^{ab}f_n\right].
\end{align}
Since both $\tau$ and $\partial_k^a$ are time-reversal odd, the Drude NOME shares the same symmetry constraints as the intrinsic NOME.
Nevertheless, this contribution becomes significant only in ultra-clean samples where $\tau$ is very large.
Finally, by combining Eq.~\eqref{eq1} with AC wavepacket formalism~\cite{Jia2024}, one can also derive a high-frequency NOME.
Remarkably, its intrinsic contribution can be both $\mathcal{T}$-even and $\mathcal{P}$-even contribution, and may therefore remain finite even in systems without SOC.

To conclude, we have developed a microscopic theory of NOME, identifying the Hermitian-connection contribution as the primary source of the effect.
This gauge-invariant formulation is suitable for first-principles implementation and quantitative comparison with experiments. Crucially, NOME obeys far fewer symmetry constraints than the linear OME, significantly expanding the candidate materials for orbital control.
Owing to their distinct scaling with the relaxation time $\tau$, we predict that the intrinsic NOME dominates in moderately disordered samples, whereas the extrinsic NOME dominates in clean samples.

\bigskip
\noindent{{\bf Acknowledgments}} ---
J.J. and Z.Q. are supported by National Natural Science Foundation of China (12488101, 12474158, and 12234017),
Innovation Program for Quantum Science and Technology (2021ZD0302800),
Anhui Initiative in Quantum Information Technologies (AHY170000).
J.W. is supported by the National Natural Science Foundation of China (12034014).
We also thank the Supercomputing Center of University of Science and Technology of China for providing the high-performance computing resources.


\begin{thebibliography}{00}
\bibitem{symmetry}
The response tensor of OME is defined as $M_a=\chi_{a;b}E_b$, which is a rank-2 pseudo-tensor.
In the 2D materials in $xy$ plane, only the orbital magnetization $M_z$ induced by in-plane electric field is relevant, thus the response tensor reduces to two independent components $\chi_{z;x}$ and $\chi_{z;y}$.
These two components are forbidden by symmetries $\mathcal{P},C_2^z, \mathcal{M}_z, C_{3,4,6}^z, S^z_{3,4,6}, C_{3,4,6}^z\mathcal{T}$ and $S_{3,4,6}^z\mathcal{T}$.
\bibitem{note1}
This constraint does not apply to orbital Chern insulators, i.e. Haldane model, where broken time-reversal ($\mathcal{T}$) symmetry allows a finite intrinsic NOME.
\end{thebibliography}
\end{document}